\documentclass{iau} 
\usepackage{graphicx}

\title[Main-belt comets: sublimation-driven activity in the asteroid belt] 
{Main-belt comets: sublimation-driven activity in the asteroid belt}

\author[Henry H.\ Hsieh]   
{Henry H.\ Hsieh$^{1,2}$
}

\affiliation{$^1$Academia Sinica Institute of Astronomy \& Astrophysics \\ 11F, Astronomy-Mathematics Building, National Taiwan University \\
No.\ 1, Sec.\ 4, Roosevelt Road \\ Taipei 10617, Taiwan \\ email: {\tt hhsieh@asiaa.sinica.edu.tw} \\[\affilskip]
$^2$Planetary Science Institute \\ 1700 East Fort Lowell Road, Suite 106 \\ Tucson, Arizona 85719, United States of America \\email: {\tt hhsieh@psi.edu}}

\pubyear{2015}
\volume{318}  
\jname{Asteroids: New Observations, New Models}
\editors{S.~R.\ Chesley, R.\ Jedicke, A.\ Morbidelli, \& D.\ Farnocchia, eds.}
\begin{document}

\maketitle

\begin{abstract}
Our knowledge of main-belt comets (MBCs), which exhibit comet-like activity likely due to the sublimation of volatile ices, yet orbit in the main asteroid belt, has increased greatly since the discovery of the first known MBC, 133P/Elst-Pizarro, in 1996, and their recognition as a new class of solar system objects after the discovery of two more MBCs in 2005.  I review work that has been done over the last 10 years to improve our understanding of these enigmatic objects, including the development of systematic discovery methods and diagnostics for distinguishing MBCs from disrupted asteroids (which exhibit comet-like activity due to physical disruptions such as impacts or rotational destabilization).  I also discuss efforts to understand the dynamical and thermal properties of these objects.
\looseness=-1
\keywords{comets: general; minor planets, asteroids; astrobiology; methods: n-body simulations; methods: numerical; techniques: photometric; techniques: spectroscopic; surveys
}
\end{abstract}

\firstsection 
\section{Introduction}

Main-belt comets (MBCs; Hsieh \& Jewitt 2006) are objects that exhibit cometary mass loss likely driven by the sublimation of volatile ice, yet occupy stable orbits in the main asteroid belt.  They comprise a subset of the active asteroids (cf.\ Jewitt 2012), which also include disrupted asteroids (cf.\ Hsieh et al.\ 2012a), which are objects that exhibit activity likely due to impacts (e.g., Jewitt et al.\ 2011; Bodewits et al.\ 2011; Ishiguro et al.\ 2011) or rotational disruption (e.g., Jewitt et al.\ 2013, 2014a; Agarwal et al.\ 2013; Drahus et al.\ 2015; Sheppard \& Trujillo 2015). Distinguishing between MBCs and disrupted asteroids is not particularly straightforward unfortunately, as gas emission (which would represent direct evidence of sublimation) is too weak to detect for any of known MBCs, despite many attempts to make such detections.  Instead, all MBCs known to date have been identified through indirect methods of inferring the presence of sublimation or by excluding all other potential dust ejection mechanisms (see Section~\ref{section:sublimation} for details).

MBCs have attracted particular attention in recent years due to the implication from their activity that they possess present-day near-surface ice (assumed to be water ice).  The current presence of near-surface ice in main-belt objects is surprising, given that water ice is unstable against sublimation at the surface temperatures of asteroids at the distance of the main belt (Hsieh et al.\ 2004).  The existence of present-day ice in the asteroid belt offers opportunities to better understand the thermal and compositional history of our solar system, place constraints on protosolar disk models, probe a potential primordial source of terrestrial water (cf.\ Morbidelli et al.\ 2000, 2012; Mottl et al.\ 2007; Owen 2008), and possibly gain insights into the conditions that allowed life to arise on Earth and that might be needed for life to arise in other extrasolar planetary systems (cf.\ Hsieh 2014).
\looseness=-1

In this proceedings paper, I briefly review work that has been done in the $\sim$10~years since the recognition of MBCs as a new cometary class, focusing particularly on recent work (i.e., in the last three years since the previous IAU General Assembly) and also focusing on MBCs rather than all active asteroids.  For a more detailed historical review of MBCs, the reader is referred to Bertini (2011), while for overviews of all types of active asteroids, the reader is referred to Jewitt (2012) and Jewitt et al.\ (2015c).
\looseness=-1

\section{Historical Highlights}

\subsection{The Discovery of the Main-Belt Comets}
\label{section:mbcdiscovery}

The prototype of the MBCs, 133P/Elst-Pizarro, was discovered as an inactive main-belt asteroid in 1979 as 1979 OW$_7$, but then was observed in 1996 exhibiting a thin dust trail aligned with its orbit plane, and was redesignated as Comet P/1996 N2 (Elst et al.\ 1996).  The prolonged emission event (several weeks or months) inferred to be responsible for the object's observed dust trail led Boehnhardt et al.\ (1998) to propose that the activity could be the result of stirring of ``icy dirt'' by a recent impact and subsequent sublimation of that ice, since an impact alone should have produced an impulsive emission event. However, no explanation was offered regarding how near-surface volatile material could have remained preserved over Gyr timescales in the warm inner solar system.
\looseness=-1

\begin{figure}[ht]
\begin{center}
 \includegraphics[width=4.5in]{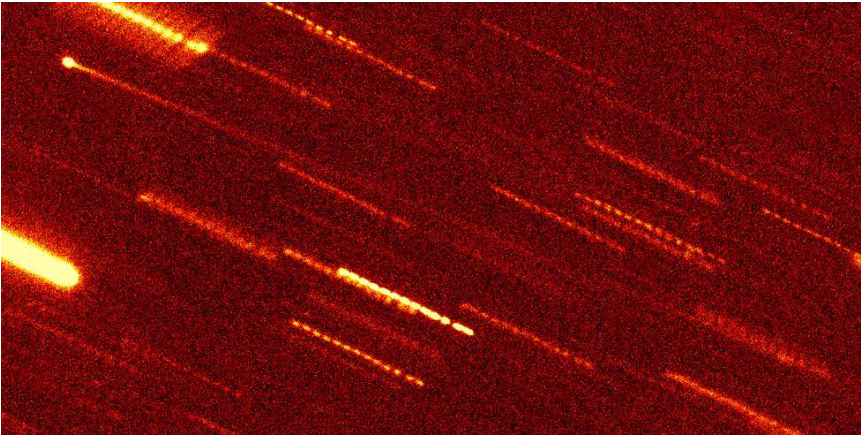} 
 \caption{Composite R-band image of 133P/Elst-Pizarro from data obtained on 2002 September 7 with the University of Hawaii 2.2 m telescope. The nucleus is at upper left with the dust trail extending across the entire field, down and to the right. The image is 2.0 arcmin by 3.5 arcmin in size, with north at the top and east to the left.
}
   \label{fig:elstpiz}
\end{center}
\end{figure}

In 2002, 133P was observed to be active again (Figure~\ref{fig:elstpiz}; Hsieh et al.\ 2004).  While emission during this active episode was once again inferred to have occurred over an extended period of time, consistent with sublimation as the driver of the activity and inconsistent with an impact, the recurrence of the activity itself provided even stronger evidence that the observed dust emission was sublimation-driven.  Repeated impact-induced dust emission events would not be expected on the same object on such a short timescale, given that such events are not observed on similar timescales on other main-belt asteroids.  On the other hand, recurrent activity near an object's perihelion is a natural consequence of sublimation-driven (i.e., cometary) dust emission, solidifying this as the most plausible explanation of 133P's observed behavior out of all of the other considered alternatives (i.e., impacts, rotational mass loss, and electrostatic dust levitation).
\looseness=-1

The finding that 133P's activity was likely cometary prompted a search for more 133P-like cometary objects in the main asteroid belt with the aim of testing two competing hypotheses for explaining 133P's origin (Hsieh 2009).  If 133P contained present-day ice because it was a highly dynamically evolved Jupiter-family comet (JFC) originally from the outer solar system that had recently evolved onto its current main-belt orbit, the apparent scarcity of dynamical pathways allowing it to do so (e.g., Fern\'andez et al.\ 2002) suggested that such objects should be rare, and that 133P could be unique.  On the other hand, if 133P was a dynamically ordinary main-belt asteroid that happened to have been able to preserve ice until the present day, other asteroids might also be icy, meaning that 133P-like objects could be common.  A survey of other main-belt asteroids could reveal more 133P-like objects exhibiting cometary activity.  In the end, this survey resulted in the 2005 discovery of activity in Themis family asteroid (118401) 1999 RE$_{70}$, since re-designated as 176P/LINEAR (Hsieh et al.\ 2006).  Along with the serendipitous discovery of 238P/Read around the same time (Read et al.\ 2005), three main-belt objects exhibiting comet-like activity were now known, all discovered from extremely limited data (133P and 238P had both been discovered by $\sim$1~m telescopes, while 176P had been discovered by an 8~m telescope as part of a deep survey of only $\sim$600 objects).  Hsieh \& Jewitt (2006) therefore concluded that a much larger undiscovered population (perhaps on the order of $\sim$100 such objects) could exist in the asteroid belt, leading to the recognition of MBCs as a new dynamical class of comets, making the asteroid belt the third known source of cometary (i.e., ice-bearing) objects in the Solar system after the Kuiper Belt and Oort Cloud.
\looseness=-1

Following the recognition of MBCs as a new cometary class in 2006, the next MBC to be discovered was 259P/Garradd (Garradd et al.\ 2008), which was also the first MBC found to have a semimajor axis interior to the 5:2 mean-motion resonance with Jupiter at 2.824~AU (all other MBCs had semimajor axes exterior to the 7:3 mean-motion resonance with Jupiter at 3.050~AU).  Besides demonstrating that MBCs could be widespread throughout the asteroid belt and not just in the outer main belt, 259P was also found to be dynamically unstable (Jewitt et al.\ 2009), forcing reconsideration of the possibility that not all MBCs may have formed at their current positions and some could have instead migrated from elsewhere in the asteroid belt, or even elsewhere in the solar system.
\looseness=-1

\subsection{Other Historical Highlights}

Other significant milestones in the study of MBCs include the discoveries of disrupted asteroids and the connection between MBCs and young asteroid families.  In 2010, dust emission associated with main-belt objects P/2010 A2 (LINEAR) and (596) Scheila were both later determined to be due to physical disruption events (impacts or rotational destabilization) rather than sublimation (Jewitt et al.\ 2010, 2011; Snodgrass et al.\ 2010; Bodewits et al.\ 2011; Ishiguro et al.\ 2011).  These discoveries provided essential reminders that sublimation cannot be assumed to be the cause of all comet-like activity in small solar system objects, and underscored the need to individually determine the most likely cause(s) of activity for each newly discovered active asteroid.  In recognition of the range of mechanisms that could potentially produce observable mass loss in asteroids, the terms ``active asteroids'' and ``disrupted asteroids'' were adopted to describe all objects with asteroid-like orbits (i.e., with a Tisserand parameter value of $T_J>3$; Vaghi 1973; Kres\'ak 1980) that exhibit visible mass loss (cf.\ Jewitt 2012), and objects with asteroid-like orbits that specifically exhibit mass loss due to impacts or rotational destabilization (cf.\ Hsieh et al.\ 2012a), respectively.
\looseness=-1

Meanwhile, at least two MBCs --- 133P and 288P/(300163) 2006 VW$_{139}$ --- have been found to be members of young ($<$10~Myr) asteroid families, where 133P has been linked to the $<$10~Myr-old Beagle family (Nesvorn\'y et al.\ 2008) and 288P has been associated with its own 7.5$\pm$0.3~Myr cluster (Novakovi\'c et al.\ 2012).  These family associations are intriguing because of thermal modeling and impact devolatilization calculations indicating that MBC-sized objects in the asteroid belt should become mostly depleted of near-surface volatile material over Gyr timescales.  Thermal modeling by Prialnik \& Rosenberg (2009) indicated that near-surface volatile material on main-belt asteroids should be severely depleted or completely exhausted over 4.6~Gyr, although Sch{\" o}rghofer (2008) showed that for certain values for obliquity and latitude, near-surface ice could in fact persist in outer main belt objects over Gyr timescales.  Meanwhile, calculations of estimated impact rates and active site depletion timescales led Hsieh (2009) to suggest that the surfaces of km-scale bodies (such as 133P) should become significantly devolatilized over Gyr timescales simply from the types of impacts hypothesized to have triggered 133P's current activity.
\looseness=-1

However, if MBCs are not primordial members of the asteroid belt and instead formed more recently in catastrophic disruptions of larger parent bodies, near-surface volatile material needs to only have persisted since the time of the fragmentation event that produced the MBC in question. Ice could not only still exist, but could also be present in much larger quantities than thermal modeling or impact rate calculations over longer timescales might suggest.  The association of two MBCs with young asteroid families appears to support this hypothesis.  While other MBCs have not yet been linked to similar young families, this does not necessarily rule out the possibility that they were also produced in recent fragmentation events.  Finding young families is dependent on having sufficiently high asteroid densities to be able to recognize overdensities in orbital element space in the population, and so it is possible that this is simply not the case for the other MBCs at the present time (but could be in the future).  Nevertheless, the collisional catastrophic disruption timescale for a 1~km object in the asteroid belt, similar to most MBCs, is on the order of a few hundreds of Myr (Cheng 2004; Bottke et al.\ 2005), suggesting that their mere existence (i.e., the fact that they were not destroyed by collisions long ago) indicates that they are likely to be recently produced.  If this is the case for all or most MBCs, it would be quite an interesting insight into the conditions necessary to form MBCs, and could suggest a method (i.e., targeted observations of members of young asteroid families) for searching for new MBCs.
\looseness=-1

\section{Evidence for Sublimation}
\label{section:sublimation}

To date, there have been no successful spectroscopic detections of any sublimation products for a MBC, despite many attempts at making such detections.  Most ground-based efforts to date have involved 8-10~m telescopes like the Keck Observatory and Gemini-North on Mauna Kea in Hawaii, Gemini-South on Cerro Pachon in Chile, the Gran Telescopio Canarias (GTC) in the Canary Islands, and the Very Large Telescope (VLT) on Cerro Paranal in Chile, and have focused on searching for emission at 3889\AA\ from the CN radical, a daughter product of HCN, which itself is commonly observed to be released during the sublimation of water ice in JFCs (cf.\ Bockel\'ee-Morvan et al.\ 2004).  These observations have found 3$\sigma$ upper limit CN production rates on the order of $Q_{\rm CN}\sim10^{21}-10^{23}$~mol~s$^{-1}$ for MBCs 133P, 259P, 288P, 313P/Gibbs, 324P/La Sagra, P/2012~T1 (PANSTARRS), and P/2013~R3 (Catalina-PANSTARRS), from which upper limit water production rates on the order of $Q_{\rm H_2O}\sim1\times10^{24}$$-$$10^{26}$~mol~s$^{-1}$ were inferred (Jewitt et al.\ 2009, 2014a, 2015b; Hsieh et al.\ 2012b, 2012c, 2013; Licandro et al.\ 2011, 2013).
\looseness=-1

The Heterodyne Instrument for the Far Infrared (HIFI) on the {\it Herschel Space Telescope} has also been used to search for line emission from 176P and P/2012 T1 at 557 GHz from the H$_2$O $1_{10}$$-$$1_{01}$ ground state rotational transition.  Again though, no emission was observed above the detection limits of the observations, where \cite{dev12} and \cite{oro13} found upper limits of $Q_{\rm H_2O}=4\times10^{25}$~mol~s$^{-1}$ for 176P and $Q_{\rm H_2O}=7.63\times10^{25}$~mol~s$^{-1}$ for P/2012 T1, respectively.  Evidence of ice on main belt objects has been obtained recently in the form of surface frost detections on (24) Themis and (90) Antiope (Rivkin \& Emery 2010; Campins et al.\ 2010; Hargrove et al.\ 2015), both members of Themis collisional family to which several other MBCs belong, and water vapor detections from dwarf planet (1) Ceres (K{\" u}ppers et al.\ 2014).  Visible dust emission has never been observed for any of these objects though, and furthermore, Ceres, Themis, and Antiope (diameters of 950~km, 200~km, and 120~km, respectively) are all much larger than the km-scale MBCs, meaning that the relevant physical conditions and processes on these three bodies are likely very different from those on the MBCs.
\looseness=-1

Despite the lack of spectroscopic confirmation of sublimation for MBCs, the action of sublimation can often be inferred indirectly.  Prolonged dust emission (as can be inferred from dust modeling; e.g., Hsieh et al.\ 2009, 2011a, 2012b, 2015b; Moreno et al.\ 2011, 2013; Jewitt et al.\ 2014b; Pozuelos et al.\ 2015) is naturally explained by sublimation, and difficult to explain as the result of an impact (cf.\ Section~\ref{section:mbcdiscovery}).  We note however that rotational disruption may be able to mimic a sustained dust emission episode through a series of mass-shedding events (cf.\ 311P/PANSTARRS; Jewitt et al.\ 2013, 2015a; Hainaut et al.\ 2014; Scheeres 2015).  On the other hand, recurrent activity near perihelion with intervening inactive periods is very difficult to explain as anything except dust emission driven by the sublimation of near-surface ice (cf.\ Hsieh et al.\ 2012a).  Repeated impact-produced active episodes on the timescale of several years for MBCs are inconsistent with the fact that similar behavior is not commonly observed for other main-belt asteroids (cf.\ Section~\ref{section:mbcdiscovery}). YORP-induced rotational mass shedding is similarly not expected to re-occur on such short timescales (e.g., Scheeres 2015).  Although recurrent activity can occasionally be confirmed by serendipitous archival observations of past activity (e.g., 313P; Hsieh et al.\ 2015b; Hui et al.\ 2015), searching for recurrent activity generally requires waiting for a full orbit period to elapse following the discovery of a new MBC candidate, making it difficult to immediately apply this criterion to classify most new discoveries.
\looseness=-1

In summary, in most cases, the case for sublimation in most MBC candidates is based on evidence ruling out other potential dust emission mechanisms, leaving sublimation as the only remaining plausible option as the main driver of activity.  It should also be noted that, in reality, multiple contributing effects may be responsible for dust emission observed for MBCs, and active asteroids in general.  For example, 133P's activity, which has been seen to repeat several times to date, is likely due to a combination of an initial triggering impact, sublimation, and rapid rotation (Hsieh et al.\ 2004, 2010; Jewitt et al.\ 2014b).  Meanwhile, although the structural failure of P/2013 R3, which disintegrated shortly following its discovery in 2013, was likely due to rotational destabilization (Jewitt et al.\ 2014a), the morphology of the dust tails observed for individual fragments indicated ongoing dust production over $>\,$2$\,-\,$3 months, suggesting that sublimation could have played a role in driving persistent dust emission for this object during its disintegration.
\looseness=-1

\section{MBC Search Efforts}

The small number of MBCs known at the present time greatly limits our ability to understand their properties as a population.  Finding more of these rare objects is crucial for advancing our knowledge of this population and for using it to trace the ice content of the inner solar system. Given the scarcity of MBCs and the transience of their activity, however, finding more is not a trivial process, requiring large amounts of observing time.  Among the three general approaches that have been used in MBC search efforts are targeted surveys, untargeted surveys, and archival searches.
\looseness=-1

Targeted surveys, such as the Hawaii Trails Project that led to the discovery of 176P (Hsieh 2009), can be designed to focus on objects with specifically selected physical and dynamical properties, and in this way, may be able to achieve relatively high discovery rates.  However, conducting individually targeted observations is time-consuming, limiting the number of objects that can be surveyed in practice, in turn limiting the total number of MBCs that can be discovered.  Furthermore, targeting objects with specific properties naturally introduces strong selection biases towards MBCs with similar properties as previously known ones, and will prevent the discovery of previously unknown types of MBCs.
\looseness=-1

On the other hand, untargeted searches, such as that conducted by the PANSTARRS1 and Palomar Transient Factory surveys (Hsieh et al.\ 2015a; Waszczak et al.\ 2013), introduce fewer selection biases than targeted surveys, and if conducted as part of larger multi-use surveys, can be much more efficient in terms of required observing effort.  The lack of strong selection biases in untargeted surveys also means that more meaningful estimates of the total size of the MBC population can be made than would be possible from a targeted survey.  Discovery rates of untargeted search will inevitably be lower than those of targeted surveys though, meaning that more objects must be surveyed to achieve each successful new discovery.  Reliable automated techniques must also be developed to search large data sets for comet candidates (e.g., Hsieh et al.\ 2012b; Waszczak et al.\ 2013), given that they will generally be too large to analyze manually.  Citizen science initiatives, where internet-based tools are used to solicit and gather large numbers of image classifications from members of the general public (e.g., Lintott et al.\ 2008) and which exploit the pattern recognition abilities of the human eye that can be difficult to replicate with automated algorithms, may also prove to be useful for analyses of these large data sets.
\looseness=-1

Finally, searches for MBCs in archival data are often able to take advantage of data sets obtained with larger telescopes (e.g., the 3.6~m Canada-France-Hawaii Telescope; Gilbert \& Wiegert 2009, 2010; Sonnett et al.\ 2011) than other large-scale survey telescopes (e.g., the 1.8~m Pan-STARRS1 telescope, or 1.5~m and 0.7~m Catalina Sky Survey telescopes), or extend over longer periods of time than any single present-day survey (e.g., Cikota et al.\ 2014).  However, the small sizes of well-controlled archival data sets mean that they may not comprise sufficiently large initial samples of asteroids for detecting rare MBCs.  For example, neither Gilbert \& Wiegert, who analyzed observations of $\sim$25,000 asteroids, nor Sonnett et al., who analyzed observations of 924 asteroids, found any MBCs, while three MBCs were found from a data set comprising $\sim$760,000 Pan-STARRS1 observations of $\sim$330,000 asteroids (Hsieh et al.\ 2015a).  Meanwhile, the variable quality of data drawn from multiple sources (e.g., the Minor Planet Center's MPCAT-OBS Observation Archive, including $\sim$75 million observations of $\sim$300,000 objects, used by Cikota et al.) makes it difficult to characterize surveys using these data very well, and may also result in false detections if data products being used (e.g., photometry) are of unreliable quality.
\looseness=-1

Archival searches do not analyze data in real time, meaning that observational confirmation or characterization of candidate comets is not as straightforward as with objects discovered in real-time surveys. If activity is expected to be recurrent though, follow-up to confirm the presence of activity can potentially be performed for a candidate MBC near the same orbital position at which activity was previously discovered (e.g., near perihelion), with the added benefit of being able to test whether activity is recurrent or not much sooner than for a MBC candidate discovered using contemporary data.  Before the 8.4~m Large Synoptic Survey Telescope becomes operational in 2022, ``surveys'' using deep archival data (e.g., from the public archives maintained by large current telescopes like the Gemini, Subaru, and Keck telescopes, and the Very Large Telescope) are a potential way to search large numbers of asteroids for activity.
Such deep archival searches could be a way to test the hypothesis proposed by Sonnett et al.\ (2011) and Hsieh et al.\ (2015a) and implied by
the discovery of faint activity for 176P by the 8.1~m Gemini North Telescope (Hsieh 2009) that weak activity could be widespread among main-belt asteroids.
\looseness=-1

\section{Dynamical Considerations}

Dynamical analysis of MBC origins is of paramount importance for assessing the utility of these objects for tracing the ice content of the modern solar system, and by extension, that of the primordial solar system.  Dynamical models of early planet migration (i.e., the so-called Nice and Grand Tack models) predict that outer solar system material could have been implanted into the current main-belt region at early times (Levison et al.\ 2009; Walsh et al.\ 2011).  While these models are by no means confirmed to accurately describe the solar system's early dynamical evolution, they make it clear that care must be taken when attempting to use MBCs to infer the properties of inner solar system ice or place constraints on conditions in the protosolar disk (cf.\ Hsieh 2014).
\looseness=-1

\begin{figure}[ht]
\begin{center}
 \includegraphics[width=5.0in]{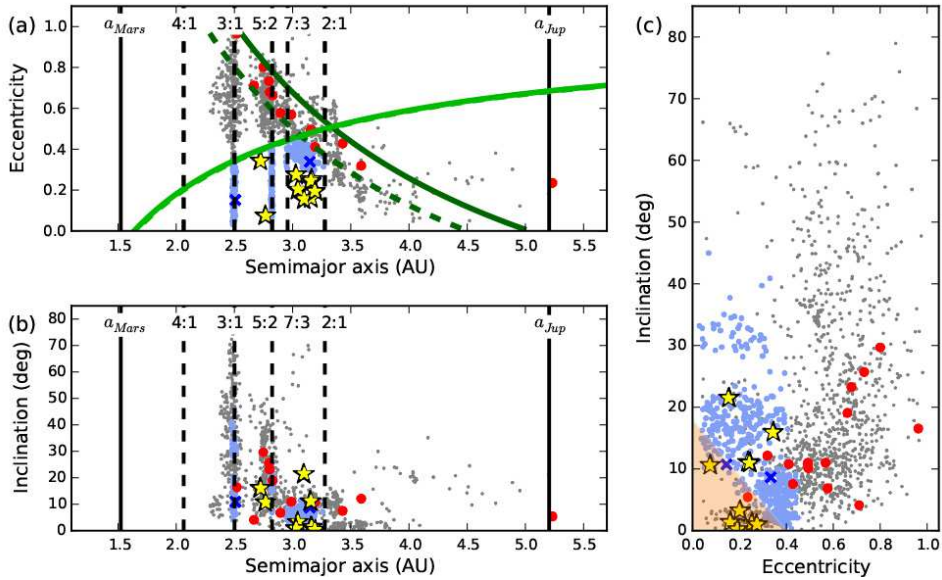} 
 \caption{Plots of (a) semimajor axis vs.\ eccentricity, (b) semimajor axis vs.\ inclination, and (c) eccentricity vs.\ inclination in 10\,000-year intervals for test particles with JFC-like initial orbital elements (red circles) that reach main-belt-like orbital elements (light blue dots) at any time over the course of the 2~Myr integration period.  Other intermediate orbital elements that are not main-belt-like are marked with gray dots, while the orbital elements of test particles with main-belt-like orbits at the end of the 2~Myr integration period are marked with blue X symbols.  Orbital elements of the known MBCs are plotted with yellow stars.  The loci of Mars-crossing orbits and Jupiter-crossing orbits are marked with light green and dark green curved solid lines, respectively, and the loci of orbits for which objects can potentially come within 1.5 Hill radii of Jupiter are marked with a dark green dashed line. In (c), the approximate region of eccentricity vs.\ inclination space into which comet-like test particles never enter is shaded in orange.  From Hsieh \& Haghighipour (2015, submitted).
 }
   \label{fig:dynamics}
\end{center}
\end{figure}

Levison \& Duncan (1994) found median dynamical lifetimes of 4.5$\times$10$^5$~years for short-period comets, but many MBCs have been found to be stable for $>$100~Myr (e.g., Haghighipour 2009; Hsieh et al.\ 2012b,c, 2013), suggesting that they may in fact have formed in situ.   There are some examples of MBCs with short ($\sim$20-30~Myr) dynamical stability timescales though (e.g., 238P and 259P; Haghighipour 2009; Jewitt et al.\ 2009), that may have formed elsewhere and were transported to their current locations.  While these objects have $T_J>3$, suggesting that they nonetheless still have asteroidal origins, purely gravitational numerical integrations of a large number of test particles by Hsieh \& Haghighipour (2015, submitted) indicate that it may be possible for objects with JFC-like initial orbital elements and $T_J<3$ to evolve onto main-belt-like orbits with $T_J>3$, at least temporarily, within 2~Myr (Figure~\ref{fig:dynamics}), apparently via repeated encounters with the terrestrial planets. These results mean that some MBCs, specifically those with high eccentricities, high inclinations, or both, could in fact be interlopers from the outer solar system, where this group includes 259P, 324P, and P/2012 T1.

This work is consistent with that of Fern\'andez et al.\ (2002) who performed purely gravitational integrations of known JFCs in an attempt to match the orbital elements of 133P.  They found that JFC 503D/Pigott eventually evolves onto a 133P-like orbit, albeit with a large inclination, a finding that has gained new significance given that high-inclination MBCs are now known.  Neither Hsieh \& Haghighipour nor Fern\'andez et al.\ (2002) were able to predict the fraction of main-belt objects that could be JFC interlopers, though Hsieh \& Haghighipour estimate that the fraction of objects with comet-like orbits occasionally reaching main-belt-like orbits could be on the order of $\sim$0.1-1\%.  Clearly, additional dynamical studies are needed to improve our understanding of the degree of JFC contamination of the main belt, and specifically of the MBC population, and the dynamical properties of those interlopers while in the main belt.
\looseness=-1

\section{MBC Activity Modulation}

\begin{figure}[ht]
\begin{center}
 \includegraphics[width=5.0in]{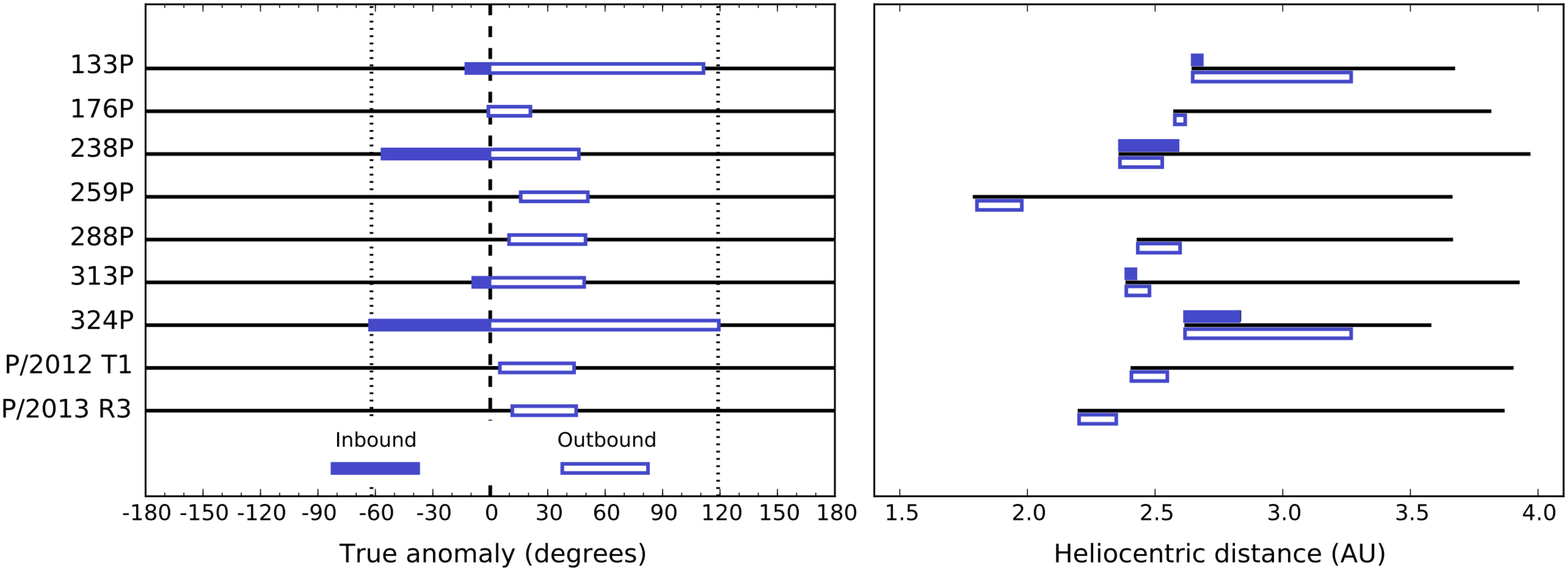} 
 \caption{Observed active ranges (between the earliest and latest observations for which activity has been reported; from Hsieh et al.\ 2015a, and references within) in terms of true anomaly (left) and heliocentric distance (right) for likely MBCs. Blue solid and outlined line segments 
indicate the inbound (pre-perihelion) and outbound (post-perihelion) portion of each object's orbit, respectively. In the left panel, perihelion is marked with a dashed vertical line, while the earliest and latest orbit positions at which activity has been observed for 324P are marked with dotted vertical lines. In the right panel, horizontal black line segments indicate the heliocentric distance range covered by the orbit of each object. From Hsieh \& Sheppard (2015).\looseness=-1
}
   \label{fig:activeranges}
\end{center}
\end{figure}

The activity of 133P was initially hypothesized to be modulated by seasonal variations in solar heating of a localized active area.  This hypothesis was motivated by the assumption that 133P's low eccentricity relative to other comets meant that its activity was unlikely to simply be controlled by solar heating intensity variations due to the changing heliocentric distance around the object's orbit.  If true, however, we would expect MBC activity to occur over a variety of true anomaly ranges, but this is not what has been observed.  Instead, all MBCs which have had their activity assessed as likely to be sublimation-driven by independent analyses have been observed to exhibit activity only near perihelion, and generally only at heliocentric distances of $<$3~AU (Fig.~\ref{fig:activeranges}).

Analysis of the orbital element distribution of the currently known MBCs reveals that activity confined to the near-perihelion portions of each object's orbit should in fact be expected, and hints toward possible conditions that may make MBC activity more likely to be discovered (Hsieh et al.\ 2015a).  Calculations of expected temperatures and sublimation rates in the asteroid belt indicate that the sublimation rate of water (in the isothermal approximation) varies by $\sim$3-4 orders of magnitude between the perihelion and aphelion distances of the known outer main-belt MBCs (i.e., with semimajor axes between the 5:2 and 2:1 mean-motion resonances with Jupiter, or 2.824~AU$\,<\,$$a$$\,<\,$3.277~AU).  As such, despite the fact that MBCs have much smaller eccentricities than other comets, the differences in sublimation rates around a typical MBC's orbit are actually large enough to explain the occurrence of sublimation-driven activity near perihelion and its absence elsewhere.
\looseness=-1

\begin{figure}[ht]
\begin{center}
 \includegraphics[width=5.3in]{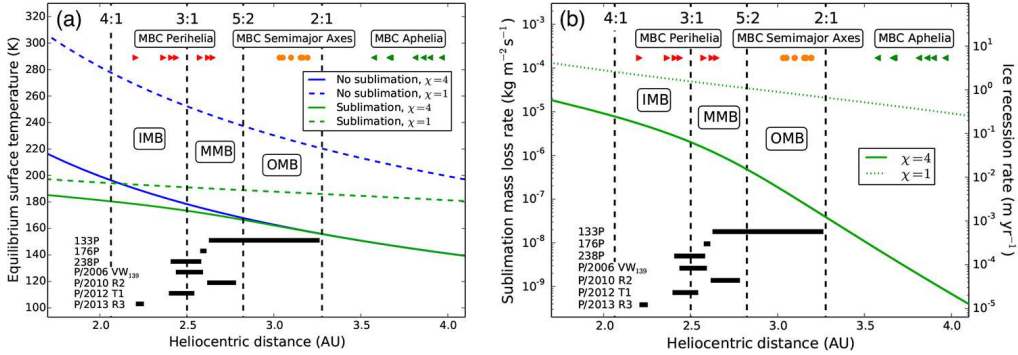} 
 \caption{(a) Equilibrium surface temperature of non-sublimating (blue lines) and sublimating (green lines) gray bodies as a function of heliocentric distance over the range of the main asteroid belt for water ice sublimation. Temperatures calculated using the isothermal approximation ($\chi$$\,=\,$4) and the subsolar approximation ($\chi$$\,=\,$1) are marked with solid and dashed lines, respectively.
(b) Mass loss rate due to water ice sublimation from a sublimating gray body as a function of heliocentric distance over the range of the main asteroid belt, where mass loss rates calculated using the isothermal approximation ($\chi$$\,=\,$4) and the subsolar approximation ($\chi$$\,=\,$1) are marked with solid and dashed green lines, respectively.  In both panels, the semimajor axis ranges of the inner, middle, and outer main belt are labeled IMB, MMB, and OMB, respectively, and the positions of the 4:1, 3:1, 5:2, and 2:1 mean-motion resonances with Jupiter that delineate the various regions of the main asteroid belt are shown with vertical dashed black lines. Also plotted are the perihelion distances (red, right-facing triangles), semimajor axis distances (orange circles), and aphelion distances (green, left-facing triangles) of the known outer main-belt MBCs, as well as the range of heliocentric distances over which they have been observed to exhibit activity (thick black horizontal lines). 
From Hsieh et al.\ (2015a).
}
   \label{fig:modulation}
\end{center}
\end{figure}

Intriguingly, the currently known outer main-belt MBCs have eccentricities of 0.15$\,<\,$$e$$\,<\,$0.30, where the eccentricity distribution of the overall outer main-belt asteroid population peaks at $e$$\,\sim\,$0.1.  This means that the known outer main-belt MBCs have perihelion distances that are systematically closer to the Sun and aphelion distances that are systematically farther from the Sun than typical members of the general outer main-belt asteroid population, and therefore will tend to reach higher temperatures at perihelion and experience lower temperatures at aphelion than the general asteroid population.  As such, they might be expected to exhibit stronger sublimation-driven activity (assuming that ice is available for sublimation) near perihelion and also experience slower depletion of their remaining volatile material elsewhere in their orbits.  Such objects might be expected to exhibit brief periods of relatively strong activity, therefore making them more likely to be discovered by present-day surveys, over a longer period of time than lower-eccentricity objects that might exhibit more uniform, but generally weaker, activity along their orbits.  As more MBCs are found in the future, it will be interesting to note whether the population continues to have higher than average eccentricities relative to the background asteroid population.

Lastly, it is interesting to note that MBCs are typically seen to be most strongly active after perihelion and also exhibit activity for longer periods of time after perihelion than prior to perihelion, rather than having their active periods centered around perihelion where solar heating is at its highest (cf.\ Fig.~\ref{fig:activeranges}).  This apparent lag in the response of sublimation strength to temperature may be due to the non-negligible time needed for a thermal wave to penetrate into each object's subsurface to reach ice (e.g., Hsieh et al.\ 2011b). This suggests that, in principle, the evolution of an object's activity strength could be used to constrain the properties (e.g., thickness, thermal inertia, porosity) of the surface layer needed to induce such a lag.  Numerical dust modeling is required to ascertain ejection times and dust production rates from observations (e.g., Hsieh et al.\ 2009; Ishiguro et al.\ 2013) since an object's observable dust cross-section does not necessarily correlate with the dust production rate at any given time. In principle though, a combination of observational monitoring and dust modeling of the evolution of each MBC's activity could be used to constrain the thermal properties of the inferred insulating material.
\looseness=-1

\section{Conclusions and Future Research}


As discussed above, there are still many issues that must be addressed before MBCs can be effectively used as probes of present-day inner solar system ice.  Many more MBCs need to be discovered (while taking care to distinguish them from disrupted asteroids) and search biases must be understood and removed to clarify their abundance and distribution in the asteroid belt. Methods must also be developed to ascertain the abundance and distribution of their dormant counterparts (i.e., icy but currently inactive asteroids), either by identifying dormant icy objects directly on large scales, or by extrapolating from the active population.  Meanwhile, further dynamical studies to ascertain the probable origins of individual MBCs, and detailed thermal models of long-term volatile depletion in icy asteroids (which may require constraints provided by observational campaigns of activity from individual objects) to clarify the relationship between an object's present-day ice content and its past ice content should help improve our ability to infer the abundance and distribution of {\it primordial} inner solar system ice from the present-day MBC population.
\looseness=-1

Perhaps still the most pressing issue facing researchers today, of course, is our continuing lack of a direct spectroscopic confirmation that sublimation is in fact the cause of MBC activity.  Without the ability to spectroscopically detect ice or gas species in MBCs with present-day facilities, we are also unable to determine the chemical and isotopic composition of the volatile material in MBCs.  Ultimately, questions about the detailed composition of MBCs will likely not be answered until we are able to send a spacecraft to one or more of these objects to conduct in situ measurements (e.g., Jones et al.\ 2015; Meech \& Castillo-Rogez 2015), making such a mission a high priority for MBC research.
\looseness=-1

Much progress has been made toward understanding MBCs in the $\sim$10 years since they were first recognized as a new cometary class, especially in the last several years, but much work still lies ahead.  Hopefully, the next several years will see even more progress in the study of these enigmatic objects and toward fulfillment of their potential to provide insights into the nature and origin of the solar system in which we live today.
\looseness=-1

\end{document}